\documentclass[aps,prb,twocolumn,superscriptaddress,showpacs,showkeys,floatfix]{revtex4}

\usepackage{graphicx,color}
\usepackage{multirow,slashbox}

\newcommand{\jwj}[1]{\textcolor{black}{#1}}

\graphicspath{{figs/}}
\begin{document}

\title{An Analytic Study of Strain Engineering the Electronic Bandgap in Single-Layer Black Phosphorus}

\author{Jin-Wu Jiang}
    \altaffiliation{Corresponding author: jiangjinwu@shu.edu.cn}
    \affiliation{Shanghai Institute of Applied Mathematics and Mechanics, Shanghai Key Laboratory of Mechanics in Energy Engineering, Shanghai University, Shanghai 200072, People's Republic of China}

\author{Harold S. Park}
    \altaffiliation{Corresponding author: parkhs@bu.edu}
    \affiliation{Department of Mechanical Engineering, Boston University, Boston, Massachusetts 02215, USA}

\date{\today}
\begin{abstract}

We present an analytic study, based on the tight-binding approximation, of strain effects on the electronic bandgap in single-layer black phosphorus.  We obtain an expression for the variation of the bandgap induced by a general strain type that includes both tension in and out of the plane and shear, and use this to determine the most efficient strain direction for different strain types, along which the strongest bandgap manipulation can be achieved.  We find that the strain direction that enables the maximum manipulation of the bandgap is not necessarily in the armchair or zigzag direction. Instead, to achieve the strongest bandgap modulation, the direction of the applied mechanical strain is dependent on the type of applied strain.

\end{abstract}
\maketitle
\pagebreak

\section{Introduction}
Strain engineering is an efficient mechanical approach to manipulating the physical properties in quasi two-dimensional nanostructures such as graphene, MoS$_2$, black phosphorus, and others.  A huge number of works have been performed to examine the effectiveness of strain in modulating the physical properties of these 2D materials, with the particularly well-known example of using strain to generate a finite electronic bandgap for graphene (for a review, see eg. Ref.~\onlinecite{CastroNAH}).

Mechanical strain has also been used to modify the physical properties in single-layer black phosphorus (SLBP).\cite{WeiQ2014apl,FeiR2014apl,OngZY2014jppc,SaB2014jpcc,KouL2014ripple,LvHY2014prb,JiangJW2014bpnpr,CaiY2015arxiv}  In particular, it has been shown in a number of previous works that mechanical strain is an effective means to tune the electronic bandgap in a wide range for SLBP. A large uniaxial strain in the direction normal to the SLBP plane can even induce a semiconductor-metal transition.\cite{RodinAS2014prl,HanX2014nl,QinGarxiv14060261,HuangGQ2014arxiv} The in-plane uniaxial strains along the armchair and zigzag directions have also been used to modify the bandgap of SLBP,\cite{PengXH2014prb,ElahiM2014prb,LiY2014jpcc} while the relative efficacy of uniaxial and biaxial strains have been comparatively studied for their effects on the electronic band structure for SLBP.\cite{FeiR2014nl,CakirD2014prb,JuW2015cpl,WangC2015jap} First-principles calculations have shown that both biaxial and uniaxial strains rotate the preferred electrical conducting direction by 90 degrees.\cite{FeiR2014nl} \jwj{The method of invariants has been applied to investigate the electronic band structure of SLBP with external fields including the strain field.\cite{VoonLCLY2015njp}}

However, in nearly all of the above works, the bandgap changes have been obtained through strains applied either in the armchair or zigzag directions, or in the direction normal to the SLBP plane. This is reasonable, because these three directions are principal directions for the \jwj{$D_{2h}$ symmetry} of the puckered configuration of the SLBP.\cite{LiP2014prb} However, a very recent study has demonstrated that the maximum in-plane Young's modulus for the SLBP is neither in the armchair direction, nor the zigzag direction. Instead, there exists a third principal direction with direction angle $\phi=0.268\pi$, along which the SLBP has the largest Young's modulus value.\cite{JiangJW2015bpyoung} Similarly, there is no guarantee that the most effective modulation of the bandgap by applying strain occurs in the armchair or zigzag direction.  

Furthermore, in most existing studies, the mechanical strain that is applied to the SLBP has been limited to either uniaxial or biaxial strain.  Hence, a natural question to ask and answer is what the optimal direction and type of mechanical strain is that results in the largest variations in the bandgap.  
A systematic analysis and understanding for the strain effect on the bandgap for a general strain type will be essential for practical strain-based manipulation of the electronic properties in SLBP.  This comprises the focus of the present work.

In this paper, using the tight-binding approximation (TBA) model, we derive an analytic formula for the strain dependence of the electronic bandgap in SLBP. We obtain an analytic expression for the direction of the applied strain, along which the strain will induce the strongest modulation in the bandgap of the SLBP. In particular, the effects from different strain types (tension, shear, and coupled tension and shear) are systematically compared.

The present paper is organized as follows. In Sec.II, we present details regarding the structure of 
SLBP. The TBA model for SLBP is introduced in Sec.III. Sec.IV~(A) is devoted to the derivation of a general analytic formula for the strain dependent bandgap, and the bandgap variations induced by different strain types are compared in Sec.IV~(B). The paper ends with a brief summary in Sec.V.

\section{Structure}

\begin{figure}[tb]
  \begin{center}
    \scalebox{0.9}[0.9]{\includegraphics[width=8cm]{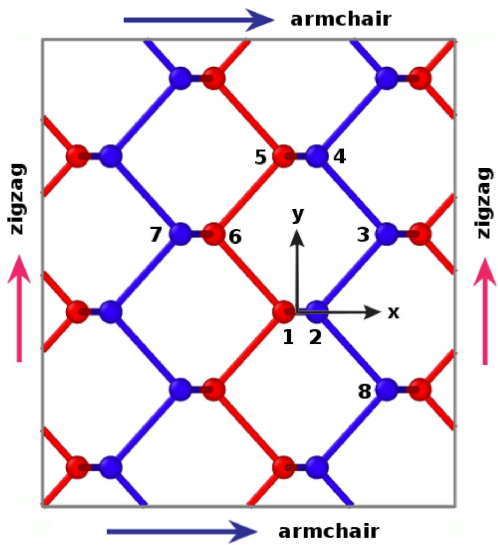}}
  \end{center}
  \caption{(Color online) SLBP structure. There are two principal directions, i.e., the armchair (blue arrows) and zigzag (red arrows) directions. Color is with respect to the atomic z-coordinate.}
  \label{fig_cfg}
\end{figure}

The atomic configuration of the SLBP is shown in Fig.~\ref{fig_cfg}. The structure parameters were measured experimentally.\cite{TakaoY1981physica} The two in-plane lattice constants are $a_1=r_{37}=4.376$~{\AA} and $a_2=r_{24}=3.314$~{\AA}, while the out-of-plane lattice constant is $a_3=10.478$~{\AA}. The origin of the Cartesian coordinate system is located in the middle of $\vec{r}_{12}$. The x-axis is in the horizontal direction and the y-axis is in the vertical direction. The z-axis is in the direction normal to the SLBP plane. There are four inequivalent atoms in the unit cell $\vec{a}_1\times\vec{a}_2$ of SLBP, which will be chosen as atoms 1, 2, 3, and 6 in this work. The coordinates of these atoms are $\vec{r}_1=(-ua_1, 0, -va_3)$, $\vec{r}_2=(ua_1, 0, va_3)$, $\vec{r}_3=(0.5a_1-ua_1, 0.5a_2, va_3)$, and $\vec{r}_6=(-0.5a_1+ua_1, 0.5a_2, -va_3)$. The two dimensionless parameters are $u=0.0806$ and $v=0.1017$. The bond lengths from the experiment are $d_1=r_{23}=r_{16}=2.2449$~{\AA} and $d_2=r_{12}=2.2340$~{\AA}, and the two angles are $\theta_{328}=0.535\pi$ and $\theta_{321}=0.567\pi$.

\section{TBA model for SLBP}
We describe now the electronic band structure for SLBP obtained using a two orbital TBA model, which is derived from a recently proposed four orbital TBA model.\cite{RudenkoAN2014prb} Specifically, it was proposed that the electronic band structure of the SLBP can be treated by a four orbital TBA model,\cite{RudenkoAN2014prb} with four hopping parameters between atom pairs (2, 3), (2, 1), (2, 6), and (3, 6) in Fig.~\ref{fig_cfg}. Among these four hopping parameters, it was shown that the electronic band structure in SLBP is determined mainly by the first two nearest-neighbor hopping parameters between atom pairs (2, 3) and (2, 1). As a consequence, we use these two leading hopping parameters to describe the electronic band structure for SLBP in the present work.

The two hopping parameters in this two orbital model are $t_{1}$ between atoms 2 and 3, and $t_{2}$ between atoms 2 and 1. For undeformed SLBP, the hopping parameter between atoms 2 and 8 ($t_3$) is the same as that between atoms 2 and 3. After the SLBP is deformed by the mechanical strain, hopping parameters $t_1$ and $t_3$ become different, so generally we have three hopping parameters in the following.

Based on the two orbital TBA model, the electronic Hamiltonian for the SLBP can be written as,
\begin{eqnarray}
H & = & \left(\begin{array}{cccc}
E_{0} & t_{2} & 0 & t_{3}+t_{1}\delta_{2}^{*}\\
t_{2} & E_{0} & t_{1}+t_{3}\delta_{2}^{*} & 0\\
0 & t_{1}+t_{3}\delta_{2} & E_{0} & t_{2}\delta_{1}\\
t_{3}+t_{1}\delta_{2} & 0 & t_{2}\delta_{1}^{*} & E_{0}
\end{array}\right)
\label{eq_ham}
\end{eqnarray}
where $\delta_{1}=e^{ik_{1}a_{1}}$ and $\delta_{2}=e^{ik_{2}a_{2}}$ are two phase factors, with $\vec{k}=k_{1}\vec{b}_{1}+k_{2}\vec{b}_{2}$ as the wave vector. Here $\vec{b}_{i}$ are two reciprocal bases defined by $\vec{b}_{i}\cdot\vec{a}_{j}=2\pi\delta_{ij}$ for $i,j=1,2$. The atomic energy level, $E_{0}$, is set to 0 in the following calculation.

The eigenvalue solution for the Hamiltonian in Eq.~(\ref{eq_ham}) gives four electronic bands for SLBP,
\begin{eqnarray}
C_{4}E^{4}+C_{2}E^{2}+C_{0} & = & 0,
\label{eq_ham_solution}
\end{eqnarray}
where the coefficients are,
\begin{eqnarray*}
C_{4} & = & 1;\\
C_{2} & = & -2\left(t_{1}^{2}+t_{2}^{2}+t_{3}^{2}+2t_{1}t_{3}\cos\Delta_{2}\right);\\
C_{0} & = & t_{2}^{4}-2t_{2}^{2}\{\cos\Delta_{1}\left[2t_{1}t_{3}+\left(t_{1}^{2}+t_{3}^{2}\right)\cos\Delta_{2}\right]\\
 &  & -\sin\Delta_{1}\sin\Delta_{2}\left(t_{1}^{2}-t_{3}^{2}\right)\}+\left(t_{1}^{2}+t_{3}^{2}+2t_{1}t_{3}\cos\Delta_{2}\right)^{2},
\end{eqnarray*}
where $\Delta_{1}=2\pi k_{1}a_{1}$ and $\Delta_{2}=2\pi k_{2}a_{2}$.

\begin{figure}[tb]
  \begin{center}
    \scalebox{1}[1]{\includegraphics[width=8cm]{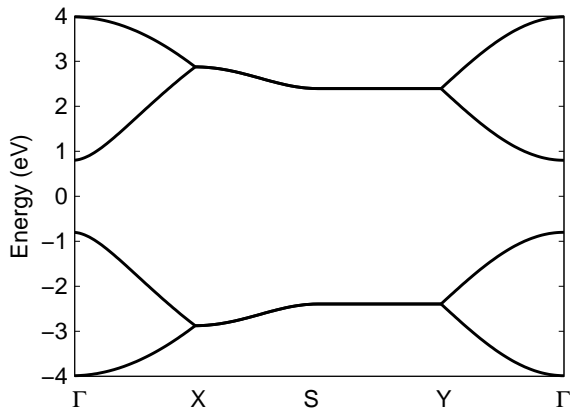}}
  \end{center}
  \caption{Electronic band structure for undeformed SLBP using the two orbital TBA model. The bandgap $\Delta E_{\rm gap}=1.60$~{eV} is reached at the $\Gamma$ point.}
  \label{fig_band}
\end{figure}

Fig.~\ref{fig_band} shows the electronic band structure from Eq.~(\ref{eq_ham_solution}) for undeformed SLBP along high symmetric lines in the first Brillouin zone. The two hopping parameters are $t_{1}^{0}=-0.797$~eV and $t_{2}^{0}=2.393$~eV. We have used the subscript 0 to denote hopping parameters in undeformed SLBP. For undeformed SLBP, the hopping parameter between atoms 2 and 8 ($t_3^0$) is the same as that between atoms 2 and 3, i.e., $t_{3}^{0}=t_1^0=-0.797$~eV. These parameters are obtained from the corresponding hopping parameters in the four orbital model by scaling them with the same factor, so that the bandgap from the two orbital TBA model agrees with that from the original four orbital model.\cite{RudenkoAN2014prb} The band structure in Fig.~\ref{fig_band} is very similar as that from the four orbital TBA model, and in particular the conductance band and the valence band are very close to the four orbital model. \jwj{The electronic band structure around the $\Gamma$ point is an even function of wave vector as restricted by the symmetry group of the SLBP.\cite{VoonLCLY2015njp}}

We focus on the bandgap modulated by mechanical strain in the SLBP. In the linear deformation regime, the direct bandgap locates at the $\Gamma$ point. The wave vector $\vec{k}=0$ at $\Gamma$ point, so the four electronic energy states are
\begin{eqnarray*}
E_{1} & = & \left(t_{1}+t_{3}\right)-t_{2}; E_{2} = -\left(t_{1}+t_{3}\right)-t_{2};\\
E_{3} & = & \left(t_{1}+t_{3}\right)+t_{2}; E_{4} = -\left(t_{1}+t_{3}\right)+t_{2}.
\end{eqnarray*}
The energy gap is,
\begin{eqnarray}
E_{\rm gap} & = & E_{3}-E_{2}=2\left(t_{1}+t_{2}+t_{3}\right).
\label{eq_gap}
\end{eqnarray}
For undeformed SLBP, we find that the bandgap $E_{\rm gap}=1.6$~eV, which agrees well with the four orbital TBA model and other first-principles calculations.\cite{RudenkoAN2014prb,YuanS2014arxiv}

\section{Strain effect on electronic bandgap}
\subsection{General formula for strain modulated bandgap}
We now consider the strain effect on the electronic bandgap of the SLBP. \jwj{The electronic bands for SLBP are composed of s and p orbitals.\cite{RudenkoAN2014prb} Moreover, the hopping parameter ($t$) between s and p orbitals depends on the bond length ($r$) as\cite{HarrisonWA1999,TangH} $t\propto\frac{1}{r^{2}}$. It has been assumed that the principal directions of the two neighboring Wannier orbitals keep their orientation along the bond vector of the two neighbor P atoms, such that the angular dependence does not play a role in the strain effect on the hopping parameter.  Instead, the strain effect is realized through changing the bond length.} Thus, the applied mechanical strain can affect electronic states (including the bandgap) through modifying the hopping parameters in the TBA model.

We consider the deformation of SLBP under a general mechanical strain in the direction with angle $\phi$.  The direction angle $\phi$ is determined starting from the x-axis, and so the armchair direction is for $\phi=0$, while the zigzag direction is for $\phi=\frac{\pi}{2}$. We perform a coordinate transformation, by rotating the x-axis in Fig.~\ref{fig_cfg} to the strain direction $\hat{e}_{\phi}=\hat{e}_x\cos\phi+\hat{e}_y\sin\phi$. The coordinates for a vector in this new coordinate system become
\begin{eqnarray}
\left(\begin{array}{c}
x_{\phi}\\
y_{\phi}\\
z_{\phi}
\end{array}\right) & = & \left(\begin{array}{ccc}
\cos\phi & \sin\phi & 0\\
-\sin\phi & \cos\phi & 0\\
0 & 0 & 1
\end{array}\right)\left(\begin{array}{c}
x\\
y\\
z
\end{array}\right),
\end{eqnarray}
where $(x,y,z)$ is the original coordinate for the vector, and the subscript $\phi$ is to denote quantities in the new coordinate system. In the new coordinate system, the coordinates are deformed by an arbitrary linear mechanical strain as
\begin{eqnarray}
\left(\begin{array}{c}
x_{\epsilon}\\
y_{\epsilon}\\
z_{\epsilon}
\end{array}\right) & = & \left(\begin{array}{ccc}
1+\epsilon_{x} & \gamma & 0\\
\gamma & 1+\epsilon_{y} & 0\\
0 & 0 & 1+\epsilon_{z}
\end{array}\right)\left(\begin{array}{c}
x_{\phi}\\
y_{\phi}\\
z_{\phi}
\end{array}\right),
\end{eqnarray}
where $\gamma$ is the shear component, while $\epsilon_x$, $\epsilon_y$, and $\epsilon_z$ are normal strains. The subscript $\epsilon$ in the coordinate is to denote quantities after deformation. We have decoupled the z component from the other two in-plane components, considering the quasi-two-dimensional nature of the SLBP structure. \jwj{The mechanical strain is applied by deforming the SLBP structure directly, which ignores the Poisson effect and does not account for subsequent structural relaxation. This treatment results in an oversimplification of the strain state, particularly as compared to those that would result in experiments, such as substrate bending and stretching, or pressing via an STM tip.  However, the error is reasonably small for linear deformation regime, as shown by comparison with prior DFT simulations in the following discussions.}

In the linear deformation regime, the bond length $r$ can be expanded as a function of all strain components, $\epsilon_{x}$, $\epsilon_{y}$, $\epsilon_{z}$, and $\gamma$ as
\begin{eqnarray}
r & = & r_{0}+\frac{\partial r}{\partial\epsilon_{x}}\epsilon_{x}+\frac{\partial r}{\partial\epsilon_{y}}\epsilon_{y}+\frac{\partial r}{\partial\epsilon_{z}}\epsilon_{z}+\frac{\partial r}{\partial\gamma}\gamma\nonumber\\
 & \equiv & r_{0}+\alpha_{x}\epsilon_{x}+\alpha_{y}\epsilon_{y}+\alpha_{z}\epsilon_{z}+\alpha_{s}\gamma,
\end{eqnarray}
where we have introduced $\alpha$ as the strain-related geometrical coefficients. Recalling the relationship between the hopping parameter and the bond length, $t\propto\frac{1}{r^{2}}$, we get the strain effect on the hopping parameter,
\begin{eqnarray}
t & = & t_{0}\left(1-\frac{2}{r_{0}}\alpha_{x}\epsilon_{x}-\frac{2}{r_{0}}\alpha_{y}\epsilon_{y}-\frac{2}{r_{0}}\alpha_{z}\epsilon_{z}-\frac{2}{r_{0}}\alpha_{s}\gamma\right).
\label{eq_t1}
\end{eqnarray}

According to Eq.~(\ref{eq_t1}), the key ingredient is to compute the strain-related geometrical coefficients $\alpha$ for each hopping parameter. For the strain $\epsilon_x$, we get the following geometrical coefficients for each hopping parameter $t_i$,
\begin{eqnarray*}
\alpha_{1}^{x} & = & \frac{\partial r_{23}}{\partial\epsilon_{x}}|_{\epsilon_{x}=0}=\frac{1}{r_{23}}x{}_{23\phi}^{2}\\
 & = & \frac{1}{d_{1}}\left[\left(0.5-2u\right)a_{1}\cos\phi+0.5a_{2}\sin\phi\right]^{2};\\
\alpha_{3}^{x} & = & \frac{\partial r_{28}}{\partial\epsilon_{x}}|_{\epsilon_{x}=0}=\frac{1}{r_{28}}x{}_{28\phi}^{2}\\
 & = & \frac{1}{d_{1}}\left[\left(0.5-2u\right)a_{1}\cos\phi-0.5a_{2}\sin\phi\right]^{2};\\
\alpha_{2}^{x} & = & \frac{\partial r_{21}}{\partial\epsilon_{x}}|_{\epsilon_{x}=0}=\frac{1}{r_{21}}x{}_{21\phi}^{2}=\frac{1}{d_{2}}\left(2ua_{1}\cos\phi\right)^{2}.
\end{eqnarray*}
Here, $\alpha_1^x$ is the coefficient corresponding to the hopping parameter $t_1$.

For the strain $\epsilon_{y}$, we obtain the following geometrical coefficients,
\begin{eqnarray*}
\alpha_{1}^{y} & = & \frac{\partial r_{23}}{\partial\epsilon_{y}}|_{\epsilon_{y}=0}=\frac{1}{r_{23}}y{}_{23\phi}^{2}\\
 & = & \frac{1}{d_{1}}\left[-\left(0.5-2u\right)a_{1}\sin\phi+0.5a_{2}\cos\phi\right]^{2};\\
\alpha_{3}^{y} & = & \frac{\partial r_{28}}{\partial\epsilon_{y}}|_{\epsilon_{y}=0}=\frac{1}{r_{28}}y{}_{28\phi}^{2}\\
 & = & \frac{1}{d_{1}}\left[\left(0.5-2u\right)a_{1}\sin\phi+0.5a_{2}\cos\phi\right]^{2};\\
\alpha_{2}^{y} & = & \frac{\partial r_{21}}{\partial\epsilon_{y}}=\frac{1}{r_{21}}y{}_{21\phi}^{2}=\frac{1}{d_{2}}\left(2ua_{1}\sin\phi\right)^{2}.
\end{eqnarray*}

For the $\epsilon_{z}$ strain, we get the following geometrical coefficients,
\begin{eqnarray*}
\alpha_{1}^{z} & = & \frac{\partial r_{23}}{\partial\epsilon_{z}}|_{\epsilon_{z}=0}=\frac{1}{r_{23}}z{}_{23\phi}^{2}=0;\\
\alpha_{3}^{z} & = & \frac{\partial r_{28}}{\partial\epsilon_{z}}|_{\epsilon_{z}=0}=\frac{1}{r_{28}}z{}_{28\phi}^{2}=0;\\
\alpha_{2}^{z} & = & \frac{\partial r_{21}}{\partial\epsilon_{z}}|_{\epsilon_{z}=0}=\frac{1}{r_{21}}z{}_{21\phi}^{2}=\frac{1}{d_{2}}\left(2va_{3}\right)^{2}.
\end{eqnarray*}

We can derive similar expressions for the geometrical coefficients, $\alpha_{s}$, corresponding to shear strain,
\begin{eqnarray*}
\alpha_{1}^{s} & = & \frac{2}{d_{1}}x{}_{23\phi}y_{23\phi}\\
 & = & \frac{2}{d_{1}}\left[\left(0.5-2u\right)a_{1}\cos\phi+0.5a_{2}\sin\phi\right]\\
 &  & \times\left[-\left(0.5-2u\right)a_{1}\sin\phi+0.5a_{2}\cos\phi\right];\\
\alpha_{3}^{s} & = & \frac{\partial r_{28}}{\partial\gamma}|_{\gamma=0}=\frac{2}{r_{28}}x{}_{28\phi}y_{28\phi}\\
 & = & \frac{2}{d_{1}}\left[\left(0.5-2u\right)a_{1}\cos\phi-0.5a_{2}\sin\phi\right]\\
 &  & \times\left[-\left(0.5-2u\right)a_{1}\sin\phi-0.5a_{2}\cos\phi\right];\\
\alpha_{2}^{s} & = & \frac{\partial r_{21}}{\partial\gamma}=\frac{2}{r_{21}}x{}_{21\phi}y_{21\phi}=\frac{2}{d_{2}}\left(-2ua_{1}\cos\phi\right)\left(2ua_{1}\sin\phi\right).
\end{eqnarray*}

Inserting these geometrical coefficients into Eq.~(\ref{eq_t1}), and using Eq.~(\ref{eq_gap}), we obtain the analytic expression for the strain dependence of the electronic bandgap,
\begin{eqnarray*}
&&E_{\rm gap}-E_{\rm gap}^0=\\
& & -4\epsilon_{x}\left[\frac{t_{1}^{0}}{d_{1}}\left(\alpha_{1}^{x}+\alpha_{3}^{x}\right)+\frac{t_{2}^{0}\alpha_{2}^{x}}{d_{2}}\right]
 -4\epsilon_{y}\left[\frac{t_{1}^{0}}{d_{1}}\left(\alpha_{1}^{y}+\alpha_{3}^{y}\right)+\frac{t_{2}^{0}\alpha_{2}^{y}}{d_{2}}\right]\\
 &  & -4\epsilon_{z}\left[\frac{t_{1}^{0}}{d_{1}}\left(\alpha_{1}^{z}+\alpha_{3}^{z}\right)+\frac{t_{2}^{0}\alpha_{2}^{z}}{d_{2}}\right]
 -4\gamma\left[\frac{t_{1}^{0}}{d_{1}}\left(\alpha_{1}^{s}+\alpha_{3}^{s}\right)+\frac{t_{2}^{0}\alpha_{2}^{s}}{d_{2}}\right].
\end{eqnarray*}
After some algebraic manipulation, we get the strain induced modification in the bandgap,
\begin{eqnarray}
\Delta E_{\rm gap} & = & e_{0}\epsilon_{z}+\left(e_{1}-2e_{2}\right)\left(\epsilon_{x}+\epsilon_{y}\right)-2e_{2}\epsilon\cos\left(2\phi+\psi\right),\nonumber\\
\label{eq_gap2}
\end{eqnarray}
where the parameters $e_{0}$, $e_{1}$ and $e_{2}$ are as follows
\begin{eqnarray}
e_{0} & = & -4\frac{t_{2}^{0}}{d_{2}^{2}}\left(2va_{3}\right)^{2}=-8.6288~{\rm eV};\\
e_{1} & = & -\frac{2t_{1}^{0}a_{2}^{2}}{d_{1}^{2}}=3.507~{\rm eV};\\
e_{2} & = & \frac{2t_{1}^{0}}{d_{1}^{2}}\left[\left[\left(0.5-2u\right)a_{1}\right]^{2}-\frac{a_{2}^{2}}{4}\right]+\frac{t_{2}^{0}\left(2ua_{1}\right)^{2}}{d_{2}^{2}}=0.411~{\rm eV}.\nonumber\\
\end{eqnarray}
We have introduced the following two quantities in the above derivation,
\begin{eqnarray}
\tan\psi & = & \frac{2\gamma}{\epsilon_{x}-\epsilon_{y}};\\
\epsilon & = & \sqrt{\left(\epsilon_{x}-\epsilon_{y}\right)^{2}+\left(2\gamma\right){}^{2}}.
\end{eqnarray}

Eq.~(\ref{eq_gap2}) shows the variation in the bandgap induced by a general strain applied in the direction with directional angle $\phi$. As can be seen from Eq.~(\ref{eq_gap2}), the variation in the bandgap depends on the strain angle $\phi$ with period $\pi$. For a given strain ratio, $\tan\psi=\frac{2\gamma}{\epsilon_{x}-\epsilon_{y}}$, the maximum (or minimum) strain effect can be achieved, if the strain is applied in the direction with angle $\phi$ satisfying
\begin{eqnarray*}
\cos\left(2\phi+\psi\right) & = & \pm1,
\end{eqnarray*}
which gives the strain direction,
\begin{eqnarray}
\phi & = & -\frac{\psi}{2}+j\frac{\pi}{2},
\label{eq_phi_general}
\end{eqnarray}
where $j$ is an integer. This means that mechanical strain can introduce the largest (smallest) modulation of the bandgap if the strain is applied in the direction described by Eq.~(\ref{eq_phi_general}). In particular, we note that, to achieve the strongest strain effect on the bandgap, there is no guarantee that the strain should be applied in the armchair or zigzag direction. Instead, the optimal strain direction is generally dependent on the type of the applied strain. 

\subsection{Comparison between different strain types}
In the above, we have derived the bandgap variation induced by a general strain in Eq.~(\ref{eq_gap2}). We have also obtained the direction for a general strain in Eq.~(\ref{eq_phi_general}), where the direction lies in the 2D plane.  This direction represents the most efficient strain direction, in that strain applied in this direction will generate the largest modulation of the bandgap.

In this section, we will determine the most efficient direction for some common strain types in SLBP. We first note that $e_{1}-2e_{2}>0$, $e_{2}>0$, and $\epsilon=\sqrt{\left(\epsilon_{x}-\epsilon_{y}\right)^{2}+\left(2\gamma\right){}^{2}}>0$ in Eq.~(\ref{eq_gap2}). It is obvious that strains $\epsilon_{x}$ and $\epsilon_{y}$ have similar effects on the bandgap, so we will discuss only one of them in some situations in the following.

(1) For uniaxial strain in the z-direction, i.e., $\epsilon_{x}=\epsilon_{y}=0$, $\gamma=0$, and $\epsilon_{z}\not=0$, we have
\begin{eqnarray}
\Delta E_{\rm gap} & = & e_{0}\epsilon_{z}.
\end{eqnarray}
We can see that the change of the bandgap is a linear function of the applied strain. This is consistent with previous first-principles
calculations.\cite{RodinAS2014prl,HanX2014nl,QinGarxiv14060261,HuangGQ2014arxiv} 

(2) For in-plane uniaxial strain, i.e., $\epsilon_{x}\not=0$, $\epsilon_{y}=0$, $\epsilon_{z}=0$ and $\gamma=0$, we have
\begin{eqnarray}
\Delta E_{\rm gap} & = & \left(e_{1}-2e_{2}\right)\epsilon_x-2e_{2}\epsilon_x\cos2\phi.
\end{eqnarray}
The most effective direction is determined by the condition that both terms on the right side have the same sign, i.e., $\cos2\phi=-1$.  This gives $\phi=\frac{\pi}{2}$, which is the zigzag direction in SLBP, and means that uniaxial strain can introduce the strongest effect on the bandgap if it is applied in the zigzag direction in SLBP.  For this uniaxial strain in the zigzag direction, the bandgap is $\Delta E_{\rm gap} = e_1 \epsilon_x$. The coefficient $e_1>0$, leading to an increase of the bandgap due to tensile strain, which is consistent with first-principles calculations.\cite{PengXH2014prb}

As another example, if we assume that the uniaxial strain $\epsilon_x$ is applied in the armchair direction ($\phi=0$), then we have $\Delta E_{\rm gap} = (e_1-4e_2) \epsilon_x$, where the coefficient $(e_1-4e_2)<e_1$.  This means that, to induce the same bandgap variation, a larger strain magnitude is needed if the uniaxial strain is applied in the armchair direction.

(3) For in-plane biaxial strain, i.e., $\epsilon_{x}=\epsilon_{y}=\epsilon$, $\epsilon_{z}=0$ and $\gamma=0$, we find,
\begin{eqnarray}
\Delta E_{\rm gap} & = & \left(e_{1}-2e_{2}\right)\left(\epsilon_{x}+\epsilon_{y}\right)=2\left(e_{1}-2e_{2}\right)\epsilon.
\end{eqnarray}
There is no preferred strain direction for biaxial strain, which is consistent with the intrinsically isotropic nature of biaxial strain.

(4) For a general in-plane strain with $\epsilon_{x}\not=\epsilon_{y}$, $\epsilon_{z}=0$ and $\gamma=0$, we find
\begin{eqnarray}
\Delta E_{\rm gap} & = & \left(e_{1}-2e_{2}\right)\left(\epsilon_{x}+\epsilon_{y}\right)-2e_{2}(\epsilon_{x}-\epsilon_{y})\cos2\phi.\nonumber\\
\end{eqnarray}
The most efficient strain direction depends on the sign of $\Delta E_{\rm gap}$. More specifically, it requires both terms on the right side to have the same sign as $\Delta E_{\rm gap}$.

For $\Delta E_{\rm gap}>0$, an effective strain application should require $\epsilon_{x}+\epsilon_{y}>0$ according to the first term on the right side. From the second term, we have $-(\epsilon_x-\epsilon_y)\cos2\phi>0$; i.e., we should have $\epsilon_x<\epsilon_y$ for $\phi=0$ or $\epsilon_x>\epsilon_y$ for $\phi=\frac{\pi}{2}$.  This indicates that the tensile strain should be applied in the two principal directions (armchair and zigzag) of SLBP, so that the bandgap can be enlarged most effectively. Furthermore, for maximum bandgap increase, the tensile strain should be larger in the zigzag direction than the armchair direction.

For $\Delta E_{\rm gap}<0$, the most effective strain application for bandgap reduction should require $\epsilon_{x}+\epsilon_{y}<0$ according to the first term on the right side. From the second term, we have $-(\epsilon_x-\epsilon_y)\cos2\phi<0$; i.e., we should have $\epsilon_x>\epsilon_y$ for $\phi=0$ or $\epsilon_x<\epsilon_y$ for $\phi=\frac{\pi}{2}$.  This indicates that the axial strain should be applied in the two principal directions (armchair and zigzag) of SLBP, so that the bandgap can be reduced most effectively. Furthermore, the compressive strain should be larger in the armchair direction than the zigzag direction to reduce the bandgap. Considering that the strain is compressive in this situation, we have larger strain magnitude in the zigzag direction than the armchair direction.

As a result, for both $\Delta E_{\rm gap}>0$ and $\Delta E_{\rm gap}<0$, strains should be applied in the two principal directions (armchair and zigzag) of SLBP. This result is consistent with recent first-principles calculations.\cite{ElahiM2014prb} Furthermore, the strain magnitude in the zigzag direction should be larger than the strain magnitude in the armchair direction to achieve the largest bandgap change.

\begin{figure}[tb]
  \begin{center}
    \scalebox{1}[1]{\includegraphics[width=8cm]{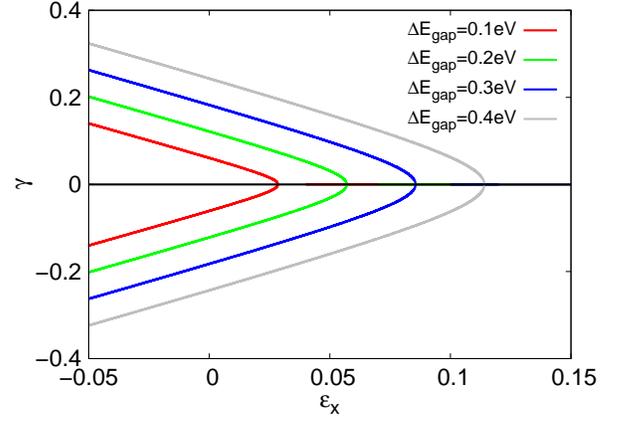}}
  \end{center}
  \caption{(Color online) The most effective approach to enlarging the bandgap by a combination of normal strain $\epsilon_x$ and shear strain $\gamma$. The direction angle for the strain is $\phi=-\frac{\psi}{2}+(2j+1)\frac{\pi}{2}$, with $\tan \psi=\frac{\gamma}{\epsilon_x}$.}
  \label{fig_contour_positive}
\end{figure}

\begin{figure}[tb]
  \begin{center}
    \scalebox{1}[1]{\includegraphics[width=8cm]{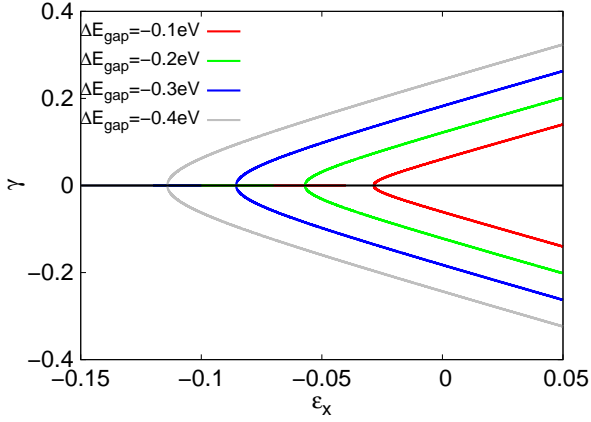}}
  \end{center}
  \caption{(Color online) The most effective approach to decreasing the bandgap by a combination of normal strain $\epsilon_x$ and shear strain $\gamma$.}
  \label{fig_contour_negative}
\end{figure}

\begin{table*}
\caption{Summary for the strain dependent bandgap variation. The last line lists the most effective direction for each strain type, along which the maximum bandgap variation can be achieved. }
\label{tab_summary}
\begin{tabular*}{\textwidth}{@{\extracolsep{\fill}}||c|c|c|c|c|c|c|c|}
\hline 
\multicolumn{2}{|c|}{strain type} & uniaxial strain & uniaxial strain & biaxial strain & general strain & shear & uniaxial strain and shear\tabularnewline
\hline 
\hline 
\multicolumn{2}{|c|}{definition} & $\epsilon_{x}=\epsilon_{y}=0$  & $\epsilon_{x}\not=0$, $\epsilon_{y}=0$  & $\epsilon_{x}=\epsilon_{y}=\epsilon$ & \multicolumn{1}{c|}{$\epsilon_{x}\not=\epsilon_{y}$ } & $\epsilon_{x}=\epsilon_{y}=0$ & $\epsilon_{x}\not=0$, $\epsilon_{y}=0$\tabularnewline
\multicolumn{2}{|c|}{} & $\epsilon_{z}\not=0$,$\gamma=0$ & $\epsilon_{z}=0$,$\gamma=0$ & $\epsilon_{z}=0$,$\gamma=0$ & $\epsilon_{z}=0$,$\gamma=0$ & $\epsilon_{z}=0$,$\gamma\not=0$ & $\epsilon_{z}=0$,$\gamma\not=0$\tabularnewline
\hline 
\multicolumn{2}{|c|}{$\Delta E_{\rm gap}$} & \multirow{2}{*}{$e_{0}\epsilon_{z}$} & $\left(e_{1}-2e_{2}\right)\epsilon_{x}$ & \multirow{2}{*}{$2\left(e_{1}-2e_{2}\right)\epsilon$} & $\left(e_{1}-2e_{2}\right)\left(\epsilon_{x}+\epsilon_{y}\right)$ & \multirow{2}{*}{$2e_{2}\gamma\sin2\phi$} & $\left(e_{1}-2e_{2}\right)\epsilon_{x}$\tabularnewline
\multicolumn{2}{|c|}{} &  & $-2e_{2}\epsilon_{x}\cos2\phi$ &  & $-2e_{2}|\epsilon_{x}-\epsilon_{y}|\cos2\phi$ &  & $-2e_{2}\sqrt{\epsilon_{x}^{2}+\gamma{}^{2}}\cos\left(2\phi+\psi\right)$\tabularnewline
\hline 
\multirow{2}{*}{$\phi_{\rm max}$} & $\Delta E_{\rm gap}>0$ & \multirow{2}{*}{N.A.} & \multirow{2}{*}{zigzag, $\phi=\frac{\pi}{2}$} & \multirow{2}{*}{arbitrary} & $\phi=\frac{\pi}{2}$, $\epsilon_x>\epsilon_y>0$ & \multirow{2}{*}{$\phi=\pm\frac{\pi}{4}$ } & $\phi=-\frac{\psi}{2}+\left(2j+1\right)\frac{\pi}{2}$\tabularnewline
\cline{2-2} \cline{6-6} \cline{8-8} 
 & $\Delta E_{\rm gap}<0$ &  &  &  & $\phi=0$, $\epsilon_y<\epsilon_x<0$ &  & $\phi=-\frac{\psi}{2}+j\pi$\tabularnewline
\hline 
\end{tabular*}
\end{table*}

(5) For pure shear strain, i.e., $\epsilon_{x}=\epsilon_{y}=\epsilon_z=0$ and $\gamma\not=0$, we find
\begin{eqnarray}
\Delta E_{\rm gap} & = & 4e_{2}\gamma\sin2\phi.
\end{eqnarray}
It is important to point out that the most effective direction for the shear strain is determined by $\sin2\phi=\pm1$, which gives $\phi=\pm\frac{\pi}{4}$, which illustrates that the most effective direction for pure shear is not in either the armchair or zigzag directions of SLBP.  Instead, a pure shear strain should be applied in the direction with $\phi=\pm\frac{\pi}{4}$, so that it can introduce the strongest effect on the bandgap for the SLBP.

(6) For strain with $\epsilon_{y}=\epsilon_z=0$, $\epsilon_x\not=0$, and $\gamma\not=0$, we simultaneously apply the uniaxial strain $\epsilon_{x}$ and the shear strain $\gamma$ to modulate the bandgap of SLBP. In this situation, we have,
\begin{eqnarray}
\Delta E_{\rm gap} & = &\left(e_{1}-2e_{2}\right)\epsilon_{x}-2e_{2}\epsilon\cos\left(2\phi+\psi\right).
\label{eq_gap_strain_shear}
\end{eqnarray}
To enlarge the bandgap, i.e., $\Delta E_{\rm gap}>0$, it can be seen from Eq.~(\ref{eq_gap_strain_shear}) that the most effective direction for applying strain is to ensure $\cos\left(2\phi+\psi\right)=-1$. This determines the angle for the strain direction,
\begin{eqnarray}
\phi & = & -\frac{\psi}{2}+(2j+1)\frac{\pi}{2},
\label{phi_psi}
\end{eqnarray}
where $j$ is an integer. Furthermore, $\epsilon_{x}$ and $\gamma$ are related to each other as,
\begin{eqnarray}
\Delta E_{\rm gap} & = & \left(e_{1}-2e_{2}\right)\epsilon_{x}+2e_{2}\sqrt{\epsilon_{x}^{2}+\left(2\gamma\right){}^{2}}.
\end{eqnarray}
Fig.~\ref{fig_contour_positive} shows this relation between $\epsilon_{x}$ and $\gamma$ for different $\Delta E_{\rm gap}$. Each curve in the figure indicates the most effective way to generate the corresponding change in the bandgap. It is clear that $\epsilon_{x}<0$ is not a good choice, because it requires larger shear strain $\gamma$. Hence, for $\Delta E_{\rm gap}>0$, the most effective way is to apply a strain with $\epsilon_{x}>0$, along with an appropriate, non-zero choice of shear strain $\gamma$. If larger $\epsilon_{x}$ is applied, then the required shear component $\gamma$ is smaller.  We note again that the strain direction ($\phi$) is determined by the actual applied strain $\epsilon_{x}$ and $\gamma$, because of the relationship between $\phi$ and $\psi$ in Eq. (\ref{phi_psi}), and because $\tan\psi=\frac{2\gamma}{\epsilon_{x}}$.

Similarly, to reduce the bandgap, i.e. $\Delta E_{\rm gap}<0$, the most effective direction for applying strain is to ensure $\cos\left(2\phi+\psi\right)=1$. This determines the angle for the strain direction,
\begin{eqnarray}
\phi & = & -\frac{\psi}{2}+2j\times\frac{\pi}{2},
\end{eqnarray}
where $j$ is an integer.  Furthermore, the strains $\epsilon_{x}$ and $\gamma$ are determined by the following relation,
\begin{eqnarray}
\Delta E_{\rm gap} & = & \left(e_{1}-2e_{2}\right)\epsilon_{x}-2e_{2}\sqrt{\epsilon_{x}^{2}+\left(2\gamma\right)^{2}}.
\end{eqnarray}
Fig.~\ref{fig_contour_negative} shows this relation between $\epsilon_{x}$ and $\gamma$ for different $\Delta E_{\rm gap}$. 

The above discussions on different strain types are summarized in Tab.~\ref{tab_summary}. From the third line in the table, uniaxial strain in the direction normal to the SLBP plane is the most effective strain type to modify the bandgap.  In other words, to generate the same bandgap variation, this strain type requires the smallest strain magnitude among all strain types that have been discussed, because it has the largest pre-coefficient magnitude, $|e_0|$.  However, the ability to apply different strain types, and combinations of strain types, is dependent on the experimental technique that is utilized.  Thus, we expect that Tab.~\ref{tab_summary} can serve as a guideline for experimentalists to choose the most appropriate strain type to manipulate the bandgap.

\jwj{We note that all discussion in this work have been based on the TBA model, which does not consider the structural relaxation and orbital hybridization effects.  This approximation is suitable for linear deformation regime, but not for nonlinear deformation with larger strains where structural relaxation and orbital hybridization occur.  Because of this, the TBA model is not able to predict certain phemonena, such as the recently reported direct to indirect transition in the bandgap of SLBP.\cite{PengXH2014prb}}

\section{conclusion}
In conclusion, we have developed an analytic model based on the tight binding approximation to elucidate strain effects on the electronic bandgap in single layer black phosphorus.  We have demonstrated that the direction along which the mechanical strain is applied is critical to achieving the maximum modulation of the bandgap.  More specifically, we have performed a detailed comparison between the effects from different strain types, and for each strain type, we present predictions for the most efficient direction for the mechanical strain as summarized in Tab.~\ref{tab_summary}. 


\textbf{Acknowledgements} The authors thank A. Rudenko for helpful communications. The work is supported by the Recruitment Program of Global Youth Experts of China and the start-up funding from Shanghai University. HSP acknowledges the support of the Mechanical Engineering department at Boston University.




%
\end{document}